\documentclass{aa}
\usepackage{graphicx}
\usepackage{txfonts}
\usepackage{lipsum}
\usepackage[normalem]{ulem}
\usepackage{subcaption}
\usepackage{lscape}
\usepackage{placeins}
\usepackage{xcolor}

\begin{document}
   \title{Zeeman-Doppler imaging of active young Suns\thanks{Based on observations made with the HARPSpol instrument on the ESO~3.6~m telescope in La Silla (Chile), under programme ID 0105.D-0436.}
   }
   \subtitle{New surface magnetic field maps for BE~Cet, HN~Peg, and $\kappa^1$~Cet}
   
   \titlerunning{ZDI of young Suns}

   \author{A.~Haris\inst{1}
   \and
   N.~Routamo\inst{1}
   \and
   T.~Willamo\inst{1}
   \and
   H.~Hilo\inst{1}
   \and
   A.~Luttinen\inst{1}
   \and
   J.~J.~Lehtinen\inst{1,2}
   \and
   O.~Kochukhov\inst{3}
   \and
   T.~Hackman\inst{1}
   \and
   I.~Ilyin\inst{4}
   \and
   M.~J.~Korpi-Lagg\inst{5}
   \and
   S.~Wedemeyer\inst{6}
   \and
   M.~Tuomi\inst{1}
        }

   \institute{
   Department of Physics, P.O.~Box 64, 00014 University of Helsinki, Finland\\
   \email{andras.haris-kiss@helsinki.fi}
   \and
   Alshain Oy, Jaakkimantie 24A, FI-02140, Espoo, Finland
   \and
   Department of Physics and Astronomy, Uppsala University, Box 524, SE-751 20 Uppsala, Sweden
   \and
   Leibniz Institute for Astrophysics Potsdam (AIP), An der Sternwarte 16, D-14482 Potsdam, Germany
   \and
   Department of Computer Science, Aalto University, PO Box 15400, FI-15400 00076 Espoo, Finland
   \and
   Rosseland Centre for Solar Physics, University of Oslo, PO Box 1029, Blindern, N-0315 Oslo, Norway
   }

   \date{Received September 30, 20XX}

  \abstract
   {Studying the magnetic activity of young solar analogues is important for understanding the dynamo mechanisms of the Sun and other late-type stars.}
   {We aimed to determine the surface magnetic field topology of four stars, analogues to the young Sun. This will contribute to long-term studies of the evolution of magnetic activity on different timescales of these targets.} 
   {We obtained high resolution spectropolarimetric measurements with the HARPSpol instrument mounted at the ESO~3.6m telescope in the epoch 2021.6, and the SOFIN instrument, mounted on the 2.56m NOT telescope in epochs 2017.4 and 2021.5. The signal to noise ratio of the observations was boosted with least squares deconvolution. The surface magnetic field maps were calculated with the Zeeman-Doppler imaging technique. The magnetic field is decomposed into spherical harmonic expansions, allowing us to quantify the contributions from the axisymmetric and non-axisymmetric components, as well as the poloidal and toroidal fields.}
   {Out of the four stars observed (BE~Cet, HN~Peg, $\kappa^1$~Cet, V1358~Ori), three (BE~Cet, HN~Peg, $\kappa^1$~Cet) show clear signs of magnetic fields. Only one star, BE~Cet shows clear evidence for the dominance of non-axisymmetric modes and the poloidal field being more dominant than the toroidal. HN~Peg also shows signs of a stronger poloidal than a toroidal field, indicating an $\alpha^2$ dynamo, where the convective turbulence effects dominate over the differential rotation. The magnetic field of V1358~Ori could not be reliably reconstructed.}
   {}


   \maketitle
   \nolinenumbers

\section{Introduction}

Magnetic fields have a defining role in the evolution of stars and their environments. They guide the angular momentum loss through magnetised stellar winds \citep{matt2012, reville2015}, thereby determining the rotational evolution of stars, and their magnetic energy levels. Magnetic fields can also enhance the emission of high energy radiation and cause energetic mass ejections, amplifying the stellar winds. These activity phenomena determine the space weather and can affect the habitability of the planets orbiting these stars \citep{see2014, airapetian2020}.

The magnetic activity of late-type stars is driven by a dynamo mechanism. The dynamo of the Sun is an $\alpha \Omega$-dynamo, meaning that it is maintained by two mechanisms, the $\Omega$- and the $\alpha$-effect \citep{parker1955,2020LRSP...17....4C}. Through the $\Omega$-effect, differential rotation winds up and amplifies the poloidal magnetic field, thereby generating a toroidal magnetic field. The $\alpha$-effect converts the toroidal back into poloidal fields through rotationally influenced convective turbulence.

In young, rapidly rotating late-type stars, differential rotation is generally reported to be weaker \citep{barnes2005}, and the $\Omega$-effect is suppressed \citep{1999A&A...344..911K}, and therefore the magnetic activity is possibly driven by other dynamos ($\alpha^2$ or $\alpha^2 \Omega$ dynamos) \citep{1999A&A...346..922K,2006A&A...446.1027C}. As these stars evolve, they become magnetically less active as they lose angular momentum through magnetic breaking. This implies a transition from the $\alpha^2$ or $\alpha^2 \Omega$ dynamo regimes to the Sun's $\alpha \Omega$-dynamo \citep{2016MNRAS.462.4442S,2017LRSP...14....4B,2018A&A...616A.160V}. The study of young late-type stars therefore has a key importance in understanding the evolution of the solar dynamo.

Zeeman-Doppler imaging \citep[ZDI;][]{semel1989, brown1991} is an effective method for reconstructing the large-scale surface magnetic field maps of late-type stars. The method uses high-resolution spectropolarimetric observations to retrieve the surface magnetic field map for each vector component; the radial, meridional and azimuthal components. Out of the full Stokes vector, typically only Stokes $IV$ is used, as the linear polarisation is often too weak to measure in the limits of a feasible exposure time.

We obtained observations of magnetically active young stars, analogues to the young Sun. Four stars (BE~Cet, HN~Peg, $\kappa^1$~Cet, and V1358~Ori) were observed with the HARPSpol instrument at the ESO~3.6m telescope in August~2021, which were supplemented by additional measurements of HN~Peg by the SOFIN spectropolarimeter mounted at the 2.56m Nordic Optical Telescope (NOT) from June~2017 and July~2021.

Our primary aim was to determine the topology of the large-scale magnetic field on the surfaces of young solar analogues and to study the contributions from the toroidal and poloidal fields, and the ratio of axisymmetry to non-axisymmetry. Our results will add to long-term studies of the targets aiming at a better understanding of the dynamo mechanisms operating in such young Suns.

\section{Observations}

\subsection{Spectropolarimetry with HARPSpol}

We collected Stokes $I$ and $V$ spectrometry with the HARPS\-pol high-resolution spectropolarimeter \citep{piskunov2011}, mounted on the ESO~3.6m telescope in La Silla, Chile. The resolution of HARPS spectra is $R \approx 110\,000$. The wavelength coverage of our spectra is from $3780$ to $6835$\,Å.
We measured four stars (BE~Cet, HN~Peg, $\kappa^1$~Cet, and V1358~Ori) over the course of eight nights in August~2021. Each HARPS\-pol exposure comprises four subexposures, as described by \citet{willamo2022}. The data were reduced with the REDUCE package \citep{piskunov2002}.

All observation sets analysed in this paper are summarised in Table~\ref{obssum}. The table presents the epochs of the observations, the phase coverage of the observation set ($f_\phi$) and the number of stellar rotational phases observed ($n_\phi$). The phase coverage was calculated assuming that one observation covers 10\% of the rotational period.

Since the wavelength region of the Ca~II H\&K lines is covered by HARPSpol spectra, we derived the chromospheric H\&K emission to bolometric flux ratio, $\log R_{\rm HK}'$. The $\log R_{\rm HK}'$ values for each stellar rotational phase are shown in Fig.\,\ref{fig:caiihk} and
are listed in Table\,\ref{tab:obs_log_harpspol} of Appendix\,\ref{app:list_of_obs}.

\begin{figure*}[t!]
    \includegraphics[width=\textwidth]{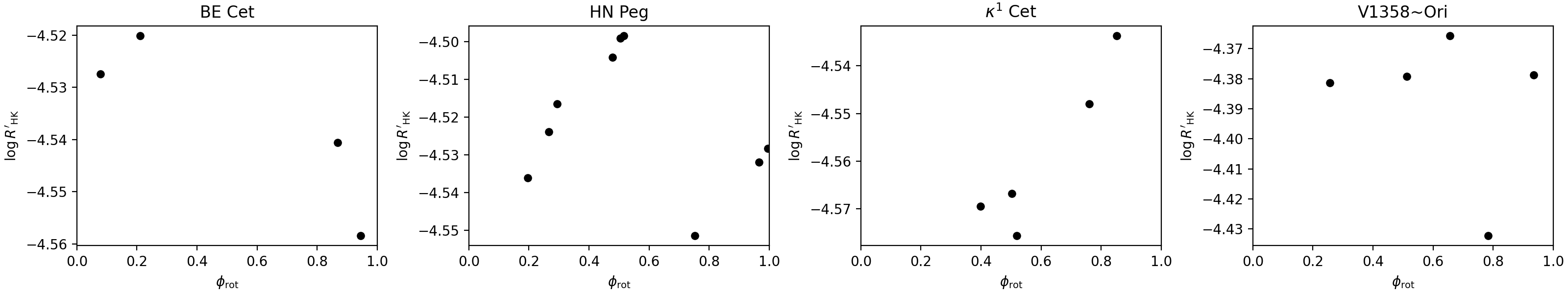}
      \caption{Chromospheric activity index $\log R_{\rm HK}'$ for HARPSpol observations.}
      \label{fig:caiihk}
\end{figure*}

\subsection{Spectropolarimetry with SOFIN}

We used the SOFIN spectropolarimeter \citep{tuominen1999} at the 2.56 m Nordic Optical Telescope (NOT) located at the Observatorio del Roque de los Muchachos on La Palma, Spain. We used a resolution of $R \approx 70\,000$, with $12$~\'echelle orders, each covering 40$-$50\,Å between 4600 and 6135\,Å.
The data were reduced with the 4A software package \citep{ilyin2000}. For a more detailed description of the instrument design and the reduction process, we refer the reader to \citet{ilyin2012}.

We measured the Stokes~$V$ spectra of HN~Peg with SOFIN over two 10-night long observing campaigns in the summers of 2017 and 2021. We obtained these spectra for five and nine stellar rotational phases in 2017 and 2021, respectively.

SOFIN is in the Cassegrain focus of the NOT and is not fibre fed. These reasons lead to a reduced wavelength stability, which may affect the null profile of the spectra, which is calculated from combining different CCD frames. This issue does not concern the Stokes~$V$ profiles, as they are derived from pairs of left- and right-handed polarisation signals, obtained on the same CCD frame \citep{hackman2026}.

\begin{table}
  \caption{Summary of the observations.}
  \centering
\begin{tabular}{llccc}
  \hline \hline

Object & Instrument & Epoch & $f_\phi$ & $n_\phi$ \\
\hline
BE~Cet         & HARPSpol & 2021.6 & 0.38 & 4  \\ 
HN~Peg         & SOFIN & 2017.4 & 0.50 & 5  \\ 
HN~Peg         & SOFIN & 2021.5 & 0.77 & 9  \\ 
HN~Peg         & HARPSpol & 2021.6  & 0.57 & 9  \\ 
$\kappa^1$~Cet & HARPSpol & 2021.6 & 0.41 & 5  \\ 
V1358~Ori      & HARPSpol & 2021.6 & 0.50 & 5  \\
\hline
\end{tabular}
\label{obssum}
\end{table}

\section{Least squares deconvolution \label{sect:lsd}}

The polarisation signatures caused by the stellar magnetic field on individual spectral lines are generally too weak to be reliably detected. To enhance the polarisation signal, we employed the least squares deconvolution method \citep[LSD,][]{donati1997}, which combines the information from thousands of spectral lines into a mean profile. We used the methodology described by \citet{donati1997} and \citet{kochukhov2010} to extract the Stokes~$I$ and $V$ LSD profiles from our spectra. We used a line mask extracted from the Vienna Atomic Line Database \citep[VALD][]{ryabchikova2015}. The resulting Stokes~$V$ profiles are shown in Fig.\,\ref{fig:stokes_v_all} of Appendix\,\ref{app:stokes_v}.

\begin{figure*}[t!]
    \includegraphics[width=\textwidth]{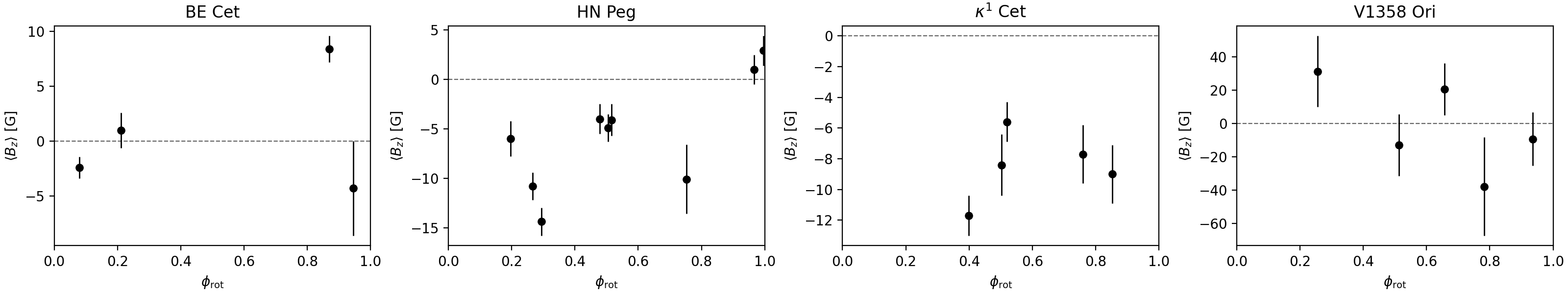}
      \caption{Mean line-of-sight component of the magnetic field for HARPSpol observations.}
      \label{fig:Bz_vs_phi}
\end{figure*}

For HARPSpol and SOFIN observations, the mean line-of-sight components of the magnetic field $\left\langle B_z \right\rangle$ were measured from the LSD Stokes $I, V$, and null profiles (only for HARPSpol observations) using the SpecPolFlow package \citep{folsom2025}, which implements the standard first-moment formalism for LSD profiles \citep{donati1997, kochukhov2010}. False alarm probabilities were calculated with SpecPolFlow, following the detection formalism of \citet{donati1992}. 
The mean line-of-sight component of the magnetic field, $\left\langle B_z \right\rangle$ for each HARPSpol observation with respect to the rotation phase is shown in Fig.\,\ref{fig:Bz_vs_phi}. 
The mean line-of-sight components of the magnetic field $\left\langle B_z \right\rangle$ from the Stokes~$V$, and null profiles (only for HARPSpol observations), and the corresponding false alarm probabilities (FAPs) are presented in Tables~\ref{tab:obs_log_harpspol} and \ref{tab:obs_log_sofin} of Appendix~\ref{app:list_of_obs}.

The FAP values for SOFIN observations are generally significantly larger than for HARPSpol observations, due to the higher noise level of the observations. Nevertheless, for both SOFIN epochs, there are at least two observations with a FAP of less than $0.0027$ (corresponding to a 3$\sigma$ detection), which should allow for a meaningful mapping of the magnetic field. We note that none of the FAP values of the HARPSpol Stokes~$V$ profiles of V1358~Ori was below the same limit, so we concluded that the mapping of the magnetic field is not possible from these spectra and omitted the object from the  later phases of the analysis.

\section{Zeeman-Doppler imaging}

The inversions of the Stokes~$V$ profiles into magnetic field maps were performed with the ZDI code \textsc{InversLSD2} developed by \citet{kochukhov2026}, based on the formulation of \citet{kochukhov2014}. The surface radial, meridional, and azimuthal magnetic field components were reconstructed using spherical harmonics, where the magnetic field components are described with spherical harmonic expansion coefficients. This allows for a straightforward calculation of the magnetic energy spectra and the determination of the energy fraction of the toroidal and poloidal components.

The number of angular orders of the expansion $\ell_{\rm max}$ was chosen based on the results of \citet{hackman2016}, who found that a higher order than five did not significantly improve the fit to the Stokes~$V$ profiles for stars with intermediate rotation velocities.
We restricted the number of angular orders of the expansion to $\ell_{\rm max} = 10$. We found that the highest angular degrees contained less than 0.1\,per~cent of the total magnetic energy of the inverted field.

Since ZDI is an inherently ill-posed inversion problem, the solution must be regularised. The magnetic field was regularised with the function \citep{kochukhov2014}:
\begin{equation}
R = \Lambda \sum_{\ell, m} \ell^2 \left ( \alpha^2_{\ell,m} + \beta^2_{\ell, m} + \gamma^2_{\ell, m} \right )\,,
\end{equation}
where $\Lambda$ is a regularisation parameter, adjusted during the inversion runs, $\alpha^2_{\ell,m}$, $\beta^2_{\ell, m}$, and $\gamma^2_{\ell, m}$ are the spherical harmonic expansion coefficients, $\ell$ is the angular degree, and $m$ is the azimuthal order.

The Stokes $V$ profiles were modelled with the Unno-Rachkovsky analytical solution of the polarised radiative transfer \citep{2004ASSL..307.....L}, assuming a triplet Zeeman splitting pattern and adjusting line width and depth to match the observed LSD Stokes $I$ spectra.
The Landé factor was taken  from the LSD profile calculations following the description of \citet{kochukhov2010}, and the resulting values are listed in Table~\ref{tab:stellar_parameters}.

Since HARPSpol has a higher resolution and a broader wavelength coverage than SOFIN, the number of spectral lines included in the LSD calculations is larger for HARPSpol ($\sim$4000) than for SOFIN ($\sim$1400). This leads to a higher boost in the signal-to-noise ratio (S/N) for HARPSpol LSD profiles. 
The differences in the amplitudes of the Stokes $V$ signals between HARPSpol and SOFIN are caused by the different lines included in the LSD calculations.

\section{Notes on individual stars \label{sect:stars}}

We adopt the ephemeris $T_0$ and the rotation period $P_{\rm rot}$ for BE~Cet and V1358~Ori from \citet{willamo2022}, and for HN~Peg and $\kappa^1$~Cet from \citet{rosen2016} and references therein. Table~\ref{tab:stellar_parameters} lists these $T_0$ and $P_{\rm rot}$ values along with other relevant parameters. The observations were phased with the ephemeris
\begin{equation}
    {\rm HJD}_{\phi = 0} = T_0 + P_{\rm rot} \times E\,.
\end{equation}
Similarly to \citet[][BE~Cet, HN~Peg, $\kappa^1$~Cet]{rosen2016} and \citet[][BE~Cet]{willamo2022}, we assume solid body rotation for all of our stars.

\begin{table*}[t!]
\centering
\caption{Stellar parameters and parameters used in the ZDI inversion. \label{tab:stellar_parameters}}
\begin{tabular}{lcccc}
\hline\hline
Parameter & BE~ Cet & HN~ Peg & $\kappa^1$ Cet & V1358~ Ori \\
\hline
Rotation period $P_{\rm rot}$ (d) & 7.676$^{(a)}$ & 4.6$^{(b)}$ & 9.2$^{(c)}$ & 1.3571$^{(d)}$ \\
Epoch $T_0$ (HJD) & 2456545.58$^{(a)}$ & 2454309.595$^{(b)}$ & 2456202.522$^{(e)}$ & 2449681.5$^{(d)}$ \\
Projected rotation velocity $v\sin i$ (km\,s$^{-1}$) & 7.0$^{(f)}$ & 10.6$^{(f)}$ & 5.2$^{(f)}$ & 41.3$^{(g)}$ \\
Inclination ($^\circ$) & 65$^{(e)}$ & 75$^{(b)}$ & 60$^{(e)}$ & 60$^{(g)}$ \\
Radius $R$ ($R_\odot$) & 1.0$^{(h)}$ & 1.04$^{(h)}$ & 0.95$^{(h)}$ & 1.05/1.08$^{(i)}$ \\
Effective temperature $T_{\rm eff}$ (K) & 5837$^{(f)}$ & 5974$^{(f)}$ & 5742$^{(f)}$ & 6032$^{(j)}$ \\
Effective Land\'e $g$ factor & 1.215$^{(k)}$ & 1.22$^{(k)}$ & 1.213$^{(k)}$ & 1.218$^{(k)}$ \\
Central wavelength $\lambda_0$ (Å), HARPS & 5126.762$^{(k)}$ & 5630.0$^{(k)*}$ & 5077.757$^{(k)}$ & 5105.357$^{(k)}$ \\
\hline
\end{tabular}
\tablefoot{
\tablefoottext{*}{For the SOFIN LSD profiles of HN~Peg, $\lambda_0=4779.17$\,Å.}
}
\begin{tablebib}
\noindent\raggedright \\
$^{(a)}$~\citet{willamo2022}.
$^{(b)}$~\citet{boro-saikia2015}.
$^{(c)}$~\citet{messina2003}.
$^{(d)}$~\citet{kriskovics2019}.
$^{(e)}$~\citet{rosen2016}.
$^{(f)}$~\citet{valenti2005}.
$^{(g)}$~\citet{hackman2016}.
$^{(h)}$~\citet{takeda2007}.
$^{(i)}$~\citet{vican2014} / \citet{zuckerman2011}.
$^{(j)}$~\citet{mcdonald2012}.
$^{(k)}$~This work.
\end{tablebib}
\end{table*}

\subsection{BE~Cet}

BE~Cet (9~Cet, HD~1835, HIP~1803) is a G2.5 type main sequence star \citep{keenan1989}, with an effective temperature of $T_{\rm eff} = 5837$\,K \citep{valenti2005} and radius of $R = 1\,R_{\odot}$ \citep{takeda2007}. Based on Gaia DR3 \citep{gaia2023}, its distance is 21.33$\pm$0.01\,pc.

Based on its membership in the Hyades cluster, its age is around 625\,Myr \citep{perryman1998}, however ages between 400\,Myr \citep{vican2012} and 850\,Myr \citep{ramirez2014} have also been estimated, based on its activity indicators $\log R'_{\rm HK} = - 4.440$ and $\log (L_{\rm X}/L_{\rm bol} ) = - 4.58$, and spectroscopic analysis, respectively. 
Estimates of the star's magnetic activity cycle range from 6.7$\pm$0.7\,yr \citep{messina2003} and 9.1$\pm$0.3\,yr \citep{brandenburg2017} to 22.0$\pm$0.9\,yr \citep{boro2018}.

Earlier ZDI inversions of BE~Cet have been published by 
\citet[][]{rosen2016}, based on HARPSpol measurements from September~2013,
\citet[][]{willamo2022}, based on HARPSpol measurements from December~2017, and 
\citet[][]{bellotti2025}, using Narval data from September~2018.

\subsection{HN~Peg}

HN~Peg (HD~206860, HIP~107350) is a G0~main sequence star with an estimated effective temperature of 5974\,K \citep{valenti2005} and a radius of 1.04\,$R_\odot$ \citep{takeda2007}. Based on Gaia DR3, the star has a distance of 18.133$\pm$0.012\,pc. It is believed to be part of the Hercules-Lyra association
which has an age of $257\pm46$\,Myr, based on gyrochronology \citep{eisenbeiss2013}. The star has a 22$\pm$9~$M_{\rm Jup}$ companion at a distance of $773 \pm 13$\,au \citep{luhman2007}. The Mount Wilson survey estimated an activity cycle period of 6.2$\pm$0.2 years \citep{baliunas1995, schroder2013}, while photometric observations showed a solar-type spot cycle of 5.5$\pm$0.3\,years \citep{messina2002}.

Two papers presented earlier ZDI inversions of HN Peg. \citet[][]{boro-saikia2015} presented six maps from between July~2007 and August~2013. These observations have also been analysed and turned into ZDI maps by \citet{rosen2016}. More recently, \citet[][]{bellotti2025} published two maps based on Narval observations from July--September~2018 and June--August~2019.

\subsection{$\kappa^1$~Cet}

$\kappa^1$~Cet (HD~20630, HIP~15457) is a G5 main sequence star, with an effective temperature of 5742\,K \citep{valenti2005} and a radius of 0.95\,$R_\odot$. It is not known to be a part of any cluster or association.
Based on Gaia DR3 data, the star's distance is 9.28$\pm$0.02\,pc. Its age has been estimated to be between 0.4 and 0.6\,Gyr \citep{ribas2010}, although ages between 0.3 \citep{lachaume1999, mamajek2008} and 0.7\,Gyr \citep{gudel1997} have also been reported. Based on a 50~year long time series of the chromospheric activity index $S_{\rm MWO}$, \citet{borosaikia2022} found two activity cycles with the periods $22.3$\,yr and $5.7$\,yr.

Three papers have presented ZDI maps of $\kappa^1$~Cet. The first, by \citet{donascimento2016} was based on Narval observations in October and November~2012. A second map was published by \citet[][]{rosen2016}, based on HARPSpol observations from September~2013.
They also re-analysed the data from \citet{donascimento2016}. The third series of maps was from \citet{borosaikia2022}, based on Narval observations from 2016, 2017, and 2018 and the re-analysis of earlier published data.

\subsection{V1358~Ori}

V1358~Ori (HD~43989, HIP~30030) is an F9 main sequence star. It is part of the Columba association, implying an age of around 30\,Myr \citep{zuckerman2011}. The star is $51.67 \pm 0.06$\,pc away based on Gaia DR3. \citet{vican2014} reported infrared excess around the star based on WISE data, and suggested the presence of a debris disk. Based on the analysis of 18~years of photometry, \citet{kriskovics2019} estimated a rotation period of $1.3571$\,d and
found a $\sim$4.4\,yr long activity cycle for the star.

Two sets of ZDI observations were turned to ZDI maps. \citet{hackman2016} and \citet{willamo2022} both observed V1358~Ori with HARPSpol in September~2013, and December~2017, respectively.

\section{Results}

\begin{figure*}[t!]
    \includegraphics[width=\textwidth]{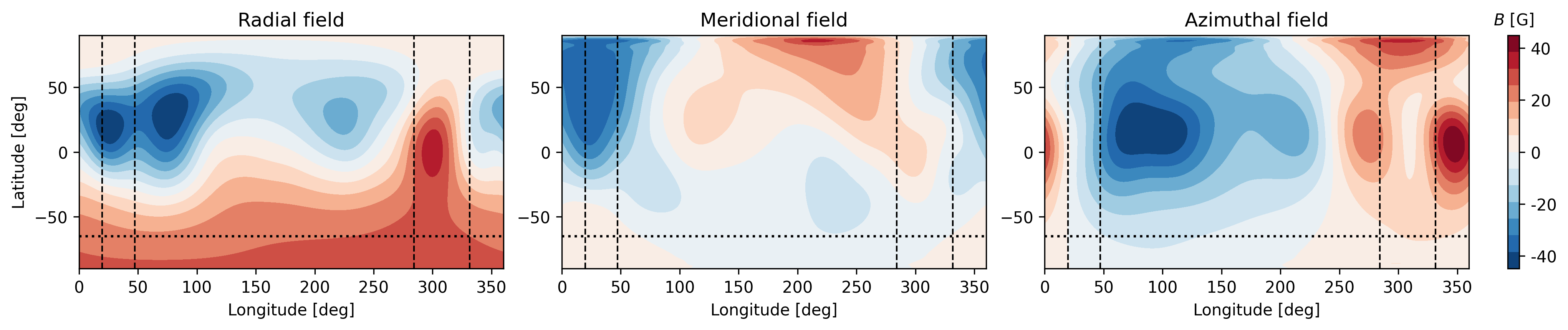}
      \caption{ZDI maps for BE~Cet from the HARPSpol measurements. Vertical lines indicate the phases of individual spectra. The horizontal line marks the lowest visible stellar latitude for the adopted inclination angle.
      }
      \label{fig:becet_full_map}
\end{figure*}

\begin{figure*}[t!]
    \includegraphics[width=\textwidth]{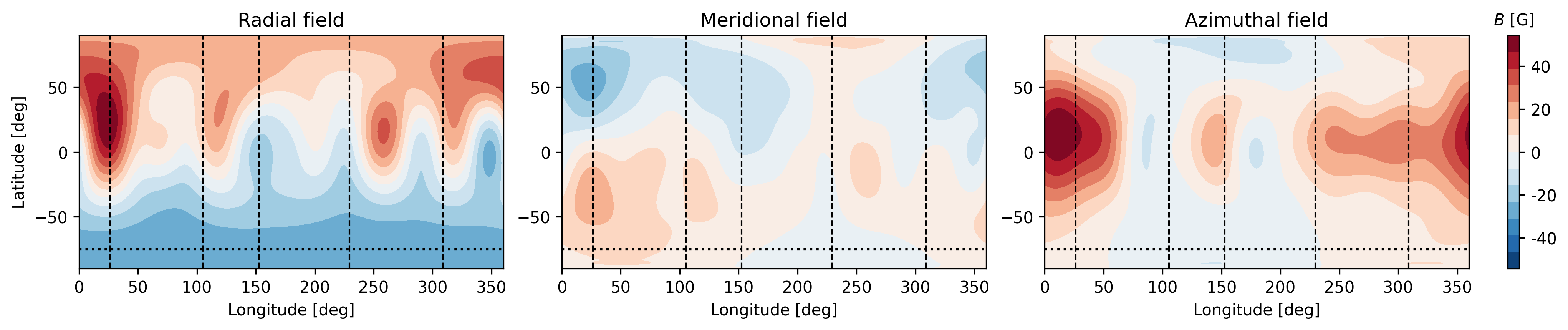}
    \includegraphics[width=\textwidth]{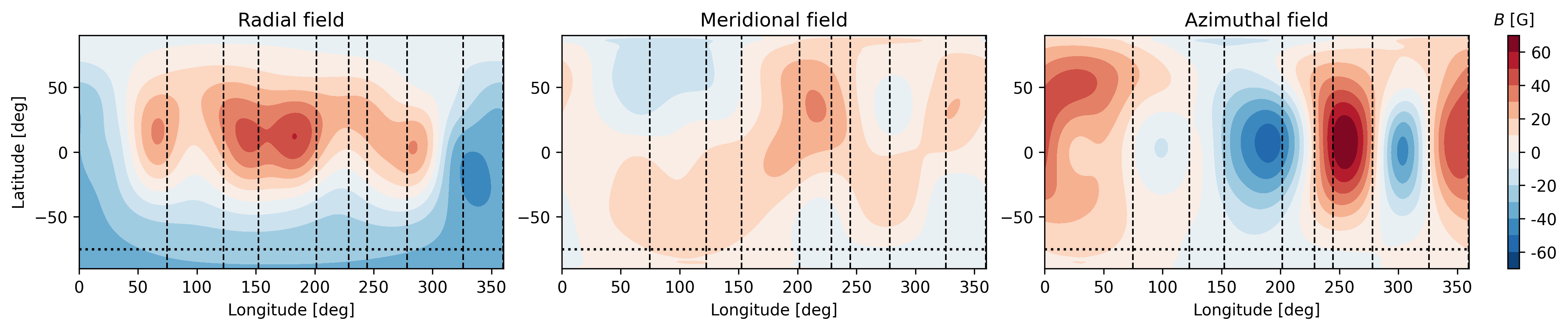}
    \includegraphics[width=\textwidth]{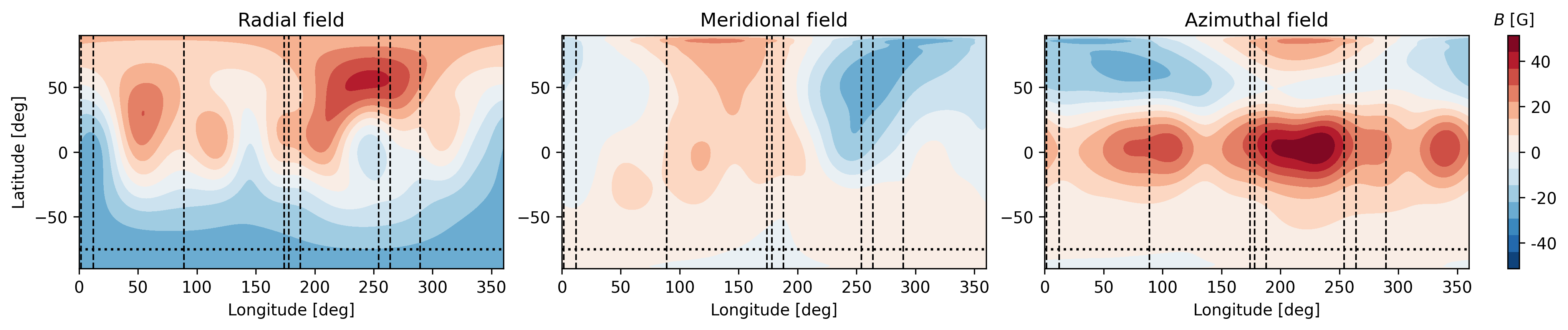}
      \caption{ZDI maps for HN~Peg. Top: based on SOFIN measurements from 2017, middle: based on SOFIN measurements from 2021, bottom: based on HARPSpol measurements from 2021.
      Markings as on Fig.~\ref{fig:becet_full_map}.
      }
      \label{fig:hnpeg_full_map_sofin2017}
\end{figure*}

\begin{figure*}[t!]
    \includegraphics[width=\textwidth]{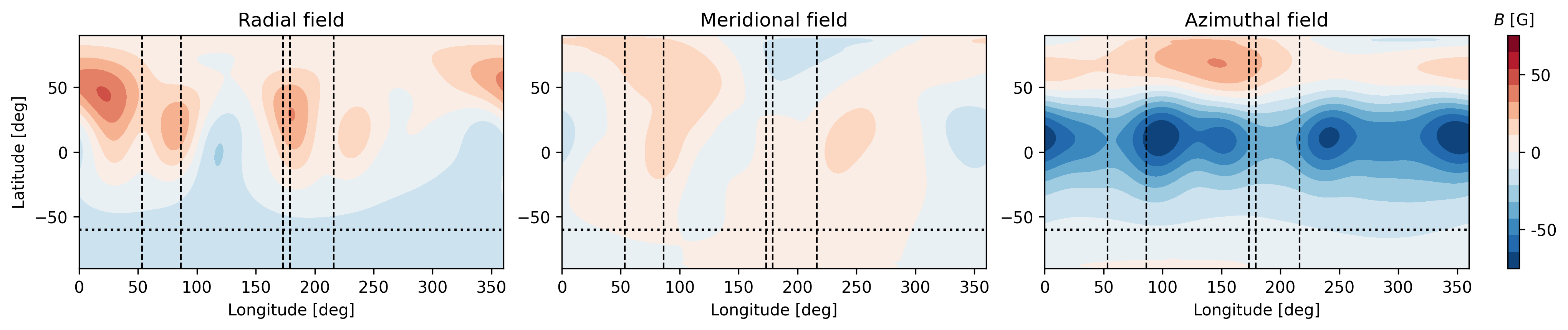}
      \caption{ZDI maps for $\kappa^1$~Cet from the HARPSpol measurements. Markings as on Fig.~\ref{fig:becet_full_map}.
      }
      \label{fig:kappa1cet_full_map}
\end{figure*}

Figs.~\ref{fig:becet_full_map}--\ref{fig:kappa1cet_full_map} show the ZDI maps for each star, for the radial, meridional, and azimuthal components of the magnetic field. Their corresponding Stokes~$V$ profiles are shown in Fig.\,\ref{fig:stokes_v_all} in Appendix\,\ref{app:stokes_v}. The strongest magnetic field is detected on $\kappa^1$~Cet, while the weakest is detected on HN~Peg (2021.6, HARPSpol, see Table~\ref{tab:app_additional_params} in Appendix~\ref{app:app_additional_params}).

\begin{figure}[t!]
    \includegraphics[width=\columnwidth]{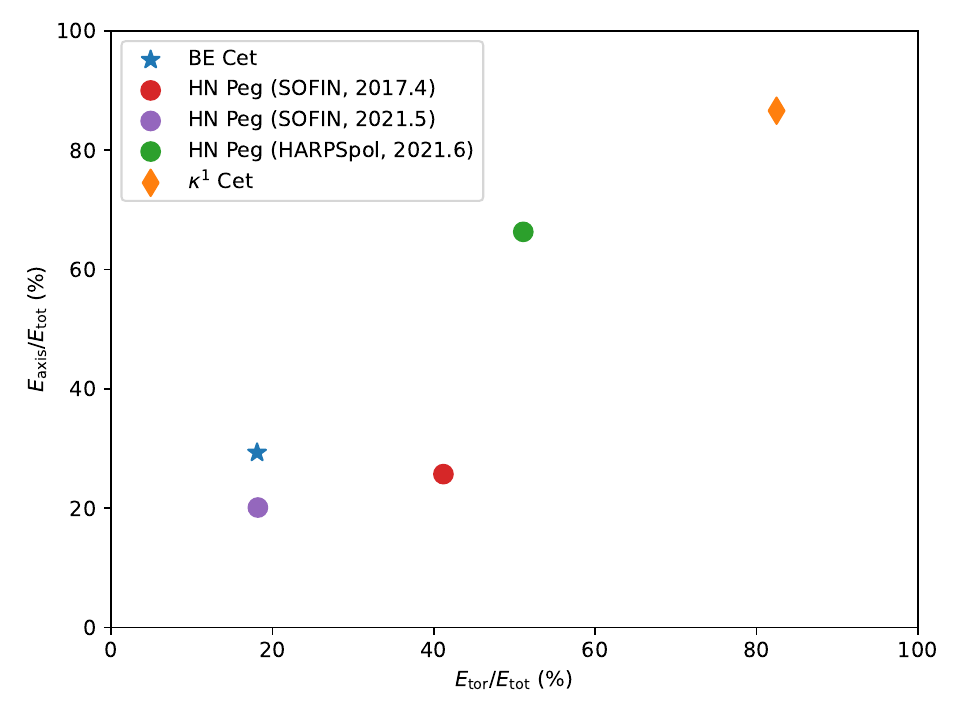}
      \caption{Fraction of the axisymmetric field energy to the total field energy with respect to the fraction of toroidal field energy to the total field energy.}
      \label{fig:eaxis_vs_etor}
\end{figure}

The spherical harmonic decomposition of the magnetic field allows us to study the distribution of magnetic energy between the poloidal and toroidal components, as well as across different angular scales. Table~\ref{tab:app_additional_params} in Appendix~\ref{app:app_additional_params} lists the mean and maximum strengths of the total magnetic field for each of our observation sets, together with the relative contributions of the poloidal and toroidal components, and the axisymmetric and non-axisymmetric components. For axisymmetry, we use the definition of \cite{fares2009}, according to which spherical harmonic modes with $|m| < \ell / 2$ are axisymmetric, while those with $|m| \geq \ell / 2$ are non-axisymmetric. This practice has been followed by \citet[e.g.][]{hackman2016} and \citet{rosen2016}. The comparison of the fraction of the axisymmetric field energy to the total field energy with respect to the fraction of toroidal field energy to the total field energy for each star in our analysis is shown in Fig.\,\ref{fig:eaxis_vs_etor}.
These results are qualitatively similar to those of \citet{rosen2016}, who studied the ZDI maps of the same stars. The biggest difference to earlier results are shown by the SOFIN observations of HN~Peg, which show lower axisymmetric and toroidal field energy fractions than those inferred for the same star from other observations.

The distribution of the total magnetic energy with respect to the angular degree of the spherical harmonics is shown in Fig.\,\ref{fig:E_vs_ell}, and listed in Table\,\ref{tab:app_fraction_E} of Appendix\,\ref{app:fraction_E}. For all of our stars, most of the magnetic field energy is concentrated in the $\ell = 1$ mode. While ZDI is more effective at retrieving the large-scale fields, some of the concentration of the field energies to lower orders may be attributed to the poor phase coverage of the data. The reason for this is that at a low phase coverage, only the largest scale features can be reconstructed with ZDI. This is supported by the notion that there is a negative correlation between the phase coverage of our observation sets and the fraction of magnetic energy in the $\ell = 1$ mode.

\begin{figure}[t!]
    \includegraphics[width=\columnwidth]{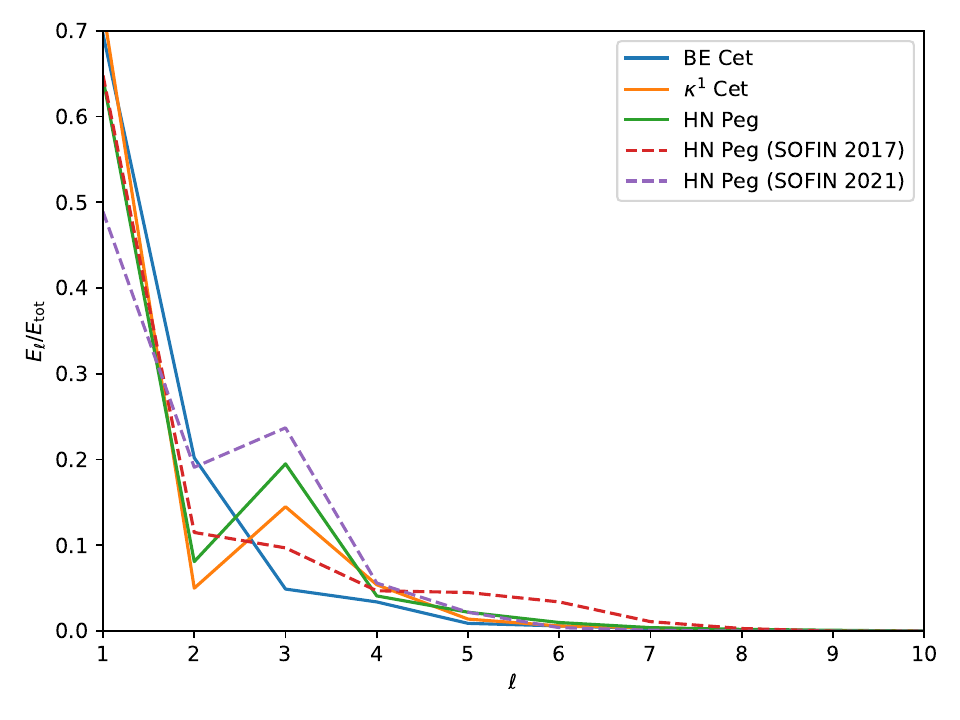}
      \caption{Distribution of the magnetic energy over the spherical harmonics. All observations are from the 2021 HARPS run, unless marked otherwise.}
      \label{fig:E_vs_ell}
\end{figure}

The maps of BE~Cet obtained by \citet{rosen2016} from 2013 show a dominant positive field, while the maps of \citet{willamo2022} and \citet{bellotti2025}, from 2017 and 2018 respectively, show a mixed polarity, with the map of \citet{bellotti2025} showing a slightly dominant negative field. We detect a weak positive field at the visible pole of BE~Cet, surrounded by a near-continuous belt of negative polarity towards lower latitudes. Due to the low phase coverage of our measurements, the latitudinal position of these features are only weakly constrained and we advise caution when interpreting this as a change in polarity.
Nearly $95\%$ of the magnetic energy is contained in $\ell=1-3$. Most of the energy is gathered in the poloidal component of the field.

For HN Peg, no major change has occurred between our observations and those published earlier. Both the radial and meridional magnetic fields exhibit a configuration similar to that seen in previous studies, with only minor local variations. The azimuthal field, by contrast, shows some changes, though it bears a resemblance to some of the earlier observations.

Some clear differences can be noted  between the magnetic field maps from 2021 based on SOFIN and HARPSpol data, although the time difference between these is only one month. This could, of course, be caused by abrupt changes in the magnetic field. However, we should take into account that the SOFIN observations have considerably lower S/N values than HARPSpol. This means that the solution will be stronger constrained by the regularisation, and, e.g., the latitudinal distribution of the radial magnetic field will not be reliable. 

For $\kappa^1$~Cet, the radial field shows a mainly positive configuration with large spots in the northern hemisphere, similar to previously published results \citep{donascimento2016, rosen2016}. Likewise, the azimuthal field has a configuration that is very similar to that found by \cite{rosen2016}; a distinct band of a negative magnetic field around the equator and a positive polar area. However, the ZDI maps of \cite{borosaikia2022} based on observations from 2016--2018 show a different configuration.

The toroidal component of the magnetic field dominates. Most of the magnetic energy is contained in $\ell=1$ and its fraction decreases towards higher $\ell$. However, a significant amount of the magnetic energy is found at $\ell=3$, consistent with the findings of \cite{rosen2016}.

For V1358~Ori, the high false alarm probability values in the Stokes~$V$ observations did not make it possible to reliably reconstruct a magnetic field for the star. For this reason, we are not presenting any ZDI inversions for V1358~Ori. A similar problem arose for the inversions of \citet{willamo2022}, based on HARPspol spectra from 2017. Out of eight phases, the magnetic field was only detected in one with a 3$\sigma$ certainty. For the HARPSpol measurements of \citet{hackman2016} taken in 2013, the magnetic field was detectable in two out of four spectra. The low amplitude of the V1358~Ori Stokes~$V$ profiles is due to the fast rotation of the star, which stretches the profiles wide and decreases their amplitudes close to the noise floor. The RMS error level of all the $\left\langle  B_z \right\rangle$ estimates of the star is rather high at 20.8\,G, which leaves considerable margin for the average longitudinal field during the observing run.

The derived dominantly non-axisymmetric and poloidal configurations of BE~Cet and HN~Peg are in good agreement with the expectation of the dynamos in stars being of $\alpha$-effect dominated type. The dominantly toroidal and axisymmetric magnetic field configuration found in $\kappa^1$~Cet, on the other hand, would be more consistent with dynamos, where also the $\Omega$-effect from differential rotation plays a prominent role.

\section{Discussion and conclusions}

We obtained high resolution spectropolarimetric observations of four analogues of the young Sun, with estimated ages between 30 and 850\,Myr. Of these four stars, we reconstructed the surface magnetic field topologies of three stars, using five sets of observations.

The configuration and strength of the magnetic field is consistent with earlier results. Comparing to earlier published observations, we found no obvious signs of polarity reversal for our stars. We note, however, that the radial field of BE~Cet shows a stronger positive polarity contribution around its pole than what is seen in its latest ZDI maps from 2018.

We note that the HARPS observations have rather low phase coverages ($f_\phi < 0.57$). All sets of observations contain less than 10 phases, with maximal phase gaps between $0.11$ (HN~Peg, SOFIN, 2021) and $0.56$ (BE~Cet, HARPS, 2021). Insufficient phase coverage is therefore expected to reduce the quality of the ZDI maps.

Since the phase gaps in the HARPSpol observations were quite large, features at low azimuthal orders $\ell$ were retrieved with much higher fidelity than smaller scale features in the magnetic field. On the other hand, in SOFIN observations, where the phases were spread rather evenly, the decline of magnetic energy towards higher $\ell$-orders is not as steep as for HARPSpol observations. However, the SOFIN observations suffered from a lower S/N. In the case of HN Peg this is a plausible explanation for the differences in reconstructed magnetic fields taken only about one month apart.

\begin{acknowledgements}
Based on observations made with the Nordic Optical Telescope, owned in collaboration by the University of Turku and Aarhus University, and operated jointly by Aarhus University, the University of Turku and the University of Oslo, representing Denmark, Finland and Norway, the University of Iceland and Stockholm University at the Observatorio del Roque de los Muchachos, La Palma, Spain, of the Instituto de Astrofisica de Canarias. The NOT data were obtained under program IDs P51-502 and P63-502.
AH acknowledges support from Otto A.~Malm Foundation.
This work has made use of the SIMBAD database, operated at CDS, Strasbourg, France. This work was supported by the Research Council of Finland (Decision 376613). 
OK acknowledges support by the Swedish Research Council (grant agreement no. 2023-03667) and the Swedish National Space Agency.
MT acknowledges the Jenny and Antti Wihuri foundation for financial support.
SW was supported by the Research Council of Norway through the Centres of Excellence scheme, project number 262622 (``Rosseland Centre for Solar Physics'').
\end{acknowledgements}

\bibliographystyle{bibtex/aa}
\bibliography{biblio}

\begin{thebibliography}{60}
\expandafter\ifx\csname natexlab\endcsname\relax\def\natexlab#1{#1}\fi

\bibitem[{{Airapetian} {et~al.}(2020){Airapetian}, {Barnes}, {Cohen}, {Collinson}, {Danchi}, {Dong}, {Del Genio}, {France}, {Garcia-Sage}, {Glocer}, {Gopalswamy}, {Grenfell}, {Gronoff}, {G{\"u}del}, {Herbst}, {Henning}, {Jackman}, {Jin}, {Johnstone}, {Kaltenegger}, {Kay}, {Kobayashi}, {Kuang}, {Li}, {Lynch}, {L{\"u}ftinger}, {Luhmann}, {Maehara}, {Mlynczak}, {Notsu}, {Osten}, {Ramirez}, {Rugheimer}, {Scheucher}, {Schlieder}, {Shibata}, {Sousa-Silva}, {Stamenkovi{\'c}}, {Strangeway}, {Usmanov}, {Vergados}, {Verkhoglyadova}, {Vidotto}, {Voytek}, {Way}, {Zank}, \& {Yamashiki}}]{airapetian2020}
{Airapetian}, V.~S., {Barnes}, R., {Cohen}, O., {et~al.} 2020, International Journal of Astrobiology, 19, 136

\bibitem[{{Baliunas} {et~al.}(1995){Baliunas}, {Donahue}, {Soon}, {Horne}, {Frazer}, {Woodard-Eklund}, {Bradford}, {Rao}, {Wilson}, {Zhang}, {Bennett}, {Briggs}, {Carroll}, {Duncan}, {Figueroa}, {Lanning}, {Misch}, {Mueller}, {Noyes}, {Poppe}, {Porter}, {Robinson}, {Russell}, {Shelton}, {Soyumer}, {Vaughan}, \& {Whitney}}]{baliunas1995}
{Baliunas}, S.~L., {Donahue}, R.~A., {Soon}, W.~H., {et~al.} 1995, \apj, 438, 269

\bibitem[{{Barnes} {et~al.}(2005){Barnes}, {Collier Cameron}, {Donati}, {James}, {Marsden}, \& {Petit}}]{barnes2005}
{Barnes}, J.~R., {Collier Cameron}, A., {Donati}, J.-F., {et~al.} 2005, \mnras, 357, L1

\bibitem[{{Bellotti} {et~al.}(2025){Bellotti}, {Petit}, {Jeffers}, {Marsden}, {Morin}, {Vidotto}, {Folsom}, {See}, \& {do Nascimento}}]{bellotti2025}
{Bellotti}, S., {Petit}, P., {Jeffers}, S.~V., {et~al.} 2025, \aap, 693, A269

\bibitem[{{Boro Saikia} {et~al.}(2015){Boro Saikia}, {Jeffers}, {Petit}, {Marsden}, {Morin}, \& {Folsom}}]{boro-saikia2015}
{Boro Saikia}, S., {Jeffers}, S.~V., {Petit}, P., {et~al.} 2015, \aap, 573, A17

\bibitem[{{Boro Saikia} {et~al.}(2022){Boro Saikia}, {L{\"u}ftinger}, {Folsom}, {Antonova}, {Alecian}, {Donati}, {Guedel}, {Hall}, {Jeffers}, {Kochukhov}, {Marsden}, {Metodieva}, {Mittag}, {Morin}, {Perdelwitz}, {Petit}, {Schmid}, \& {Vidotto}}]{borosaikia2022}
{Boro Saikia}, S., {L{\"u}ftinger}, T., {Folsom}, C.~P., {et~al.} 2022, \aap, 658, A16

\bibitem[{{Boro Saikia} {et~al.}(2018){Boro Saikia}, {Marvin}, {Jeffers}, {Reiners}, {Cameron}, {Marsden}, {Petit}, {Warnecke}, \& {Yadav}}]{boro2018}
{Boro Saikia}, S., {Marvin}, C.~J., {Jeffers}, S.~V., {et~al.} 2018, \aap, 616, A108

\bibitem[{{Brandenburg} {et~al.}(2017){Brandenburg}, {Mathur}, \& {Metcalfe}}]{brandenburg2017}
{Brandenburg}, A., {Mathur}, S., \& {Metcalfe}, T.~S. 2017, \apj, 845, 79

\bibitem[{{Brown} {et~al.}(1991){Brown}, {Donati}, {Rees}, \& {Semel}}]{brown1991}
{Brown}, S.~F., {Donati}, J.-F., {Rees}, D.~E., \& {Semel}, M. 1991, \aap, 250, 463

\bibitem[{{Brun} \& {Browning}(2017)}]{2017LRSP...14....4B}
{Brun}, A.~S. \& {Browning}, M.~K. 2017, Living Reviews in Solar Physics, 14, 4

\bibitem[{{Chabrier} \& {K{\"u}ker}(2006)}]{2006A&A...446.1027C}
{Chabrier}, G. \& {K{\"u}ker}, M. 2006, \aap, 446, 1027

\bibitem[{{Charbonneau}(2020)}]{2020LRSP...17....4C}
{Charbonneau}, P. 2020, Living Reviews in Solar Physics, 17, 4

\bibitem[{{do Nascimento} {et~al.}(2016){do Nascimento}, {Vidotto}, {Petit}, {Folsom}, {Castro}, {Marsden}, {Morin}, {Porto de Mello}, {Meibom}, {Jeffers}, {Guinan}, \& {Ribas}}]{donascimento2016}
{do Nascimento}, Jr., J.-D., {Vidotto}, A.~A., {Petit}, P., {et~al.} 2016, \apjl, 820, L15

\bibitem[{{Donati} {et~al.}(1997){Donati}, {Semel}, {Carter}, {Rees}, \& {Collier Cameron}}]{donati1997}
{Donati}, J.-F., {Semel}, M., {Carter}, B.~D., {Rees}, D.~E., \& {Collier Cameron}, A. 1997, \mnras, 291, 658

\bibitem[{{Donati} {et~al.}(1992){Donati}, {Semel}, \& {Rees}}]{donati1992}
{Donati}, J.-F., {Semel}, M., \& {Rees}, D.~E. 1992, \aap, 265, 669

\bibitem[{{Eisenbeiss} {et~al.}(2013){Eisenbeiss}, {Ammler-von Eiff}, {Roell}, {Mugrauer}, {Adam}, {Neuh{\"a}user}, {Schmidt}, \& {Bedalov}}]{eisenbeiss2013}
{Eisenbeiss}, T., {Ammler-von Eiff}, M., {Roell}, T., {et~al.} 2013, \aap, 556, A53

\bibitem[{{Fares} {et~al.}(2009){Fares}, {Donati}, {Moutou}, {Bohlender}, {Catala}, {Deleuil}, {Shkolnik}, {Collier Cameron}, {Jardine}, \& {Walker}}]{fares2009}
{Fares}, R., {Donati}, J.-F., {Moutou}, C., {et~al.} 2009, \mnras, 398, 1383

\bibitem[{{Folsom} {et~al.}(2025){Folsom}, {Erba}, {Petit}, {Seadrow}, {Stanley}, {Natan}, {Zaire}, {Oksala}, {Villadiego Forero}, {Moore}, \& {Catalan Olais}}]{folsom2025}
{Folsom}, C., {Erba}, C., {Petit}, V., {et~al.} 2025, The Journal of Open Source Software, 10, 7891

\bibitem[{{Gaia Collaboration} {et~al.}(2023){Gaia Collaboration}, {Vallenari}, {Brown}, {Prusti}, {de Bruijne}, {Arenou}, {Babusiaux}, {Biermann}, {Creevey}, {Ducourant}, {Evans}, {Eyer}, {Guerra}, {Hutton}, {Jordi}, {Klioner}, {Lammers}, {Lindegren}, {Luri}, {Mignard}, {Panem}, {Pourbaix}, {Randich}, {Sartoretti}, {Soubiran}, {Tanga}, {Walton}, {Bailer-Jones}, {Bastian}, {Drimmel}, {Jansen}, {Katz}, {Lattanzi}, {van Leeuwen}, {Bakker}, {Cacciari}, {Casta{\~n}eda}, {De Angeli}, {Fabricius}, {Fouesneau}, {Fr{\'e}mat}, {Galluccio}, {Guerrier}, {Heiter}, {Masana}, {Messineo}, {Mowlavi}, {Nicolas}, {Nienartowicz}, {Pailler}, {Panuzzo}, {Riclet}, {Roux}, {Seabroke}, {Sordo}, {Th{\'e}venin}, {Gracia-Abril}, {Portell}, {Teyssier}, {Altmann}, {Andrae}, {Audard}, {Bellas-Velidis}, {Benson}, {Berthier}, {Blomme}, {Burgess}, {Busonero}, {Busso}, {C{\'a}novas}, {Carry}, {Cellino}, {Cheek}, {Clementini}, {Damerdji}, {Davidson}, {de Teodoro}, {Nu{\~n}ez Campos}, {Delchambre}, {Dell'Oro}, {Esquej},
  {Fern{\'a}ndez-Hern{\'a}ndez}, {Fraile}, {Garabato}, {Garc{\'\i}a-Lario}, {Gosset}, {Haigron}, {Halbwachs}, {Hambly}, {Harrison}, {Hern{\'a}ndez}, {Hestroffer}, {Hodgkin}, {Holl}, {Jan{\ss}en}, {Jevardat de Fombelle}, {Jordan}, {Krone-Martins}, {Lanzafame}, {L{\"o}ffler}, {Marchal}, {Marrese}, {Moitinho}, {Muinonen}, {Osborne}, {Pancino}, {Pauwels}, {Recio-Blanco}, {Reyl{\'e}}, {Riello}, {Rimoldini}, {Roegiers}, {Rybizki}, {Sarro}, {Siopis}, {Smith}, {Sozzetti}, {Utrilla}, {van Leeuwen}, {Abbas}, {{\'A}brah{\'a}m}, {Abreu Aramburu}, {Aerts}, {Aguado}, {Ajaj}, {Aldea-Montero}, {Altavilla}, {{\'A}lvarez}, {Alves}, {Anders}, {Anderson}, {Anglada Varela}, {Antoja}, {Baines}, {Baker}, {Balaguer-N{\'u}{\~n}ez}, {Balbinot}, {Balog}, {Barache}, {Barbato}, {Barros}, {Barstow}, {Bartolom{\'e}}, {Bassilana}, {Bauchet}, {Becciani}, {Bellazzini}, {Berihuete}, {Bernet}, {Bertone}, {Bianchi}, {Binnenfeld}, {Blanco-Cuaresma}, {Blazere}, {Boch}, {Bombrun}, {Bossini}, {Bouquillon}, {Bragaglia}, {Bramante}, {Breedt},
  {Bressan}, {Brouillet}, {Brugaletta}, {Bucciarelli}, {Burlacu}, {Butkevich}, {Buzzi}, {Caffau}, {Cancelliere}, {Cantat-Gaudin}, {Carballo}, {Carlucci}, {Carnerero}, {Carrasco}, {Casamiquela}, {Castellani}, {Castro-Ginard}, {Chaoul}, {Charlot}, {Chemin}, {Chiaramida}, {Chiavassa}, {Chornay}, {Comoretto}, {Contursi}, {Cooper}, {Cornez}, {Cowell}, {Crifo}, {Cropper}, {Crosta}, {Crowley}, {Dafonte}, {Dapergolas}, {David}, {David}, {de Laverny}, {De Luise}, \& {De March}}]{gaia2023}
{Gaia Collaboration}, {Vallenari}, A., {Brown}, A.~G.~A., {et~al.} 2023, \aap, 674, A1

\bibitem[{{G{\"u}del} {et~al.}(1997){G{\"u}del}, {Guinan}, \& {Skinner}}]{gudel1997}
{G{\"u}del}, M., {Guinan}, E.~F., \& {Skinner}, S.~L. 1997, \apj, 483, 947

\bibitem[{{Hackman} {et~al.}(2016){Hackman}, {Lehtinen}, {Ros{\'e}n}, {Kochukhov}, \& {K{\"a}pyl{\"a}}}]{hackman2016}
{Hackman}, T., {Lehtinen}, J., {Ros{\'e}n}, L., {Kochukhov}, O., \& {K{\"a}pyl{\"a}}, M.~J. 2016, \aap, 587, A28

\bibitem[{{Hackman} {et~al.}(2026){Hackman}, {Lehtinen}, {Kochukhov}, {Pirttikangas}, {Ilyin}, {Haris}, {Korpi-Lagg}, {Wedemeyer}, \& {Tuomi}}]{hackman2026}
{Hackman}, T., {Lehtinen}, J.~J., {Kochukhov}, O., {et~al.} 2026, submitted

\bibitem[{{Ilyin}(2012)}]{ilyin2012}
{Ilyin}, I. 2012, Astronomische Nachrichten, 333, 213

\bibitem[{{Ilyin}(2000)}]{ilyin2000}
{Ilyin}, I.~V. 2000, PhD thesis, University of Oulu, Division of Astronomy

\bibitem[{{Keenan} \& {McNeil}(1989)}]{keenan1989}
{Keenan}, P.~C. \& {McNeil}, R.~C. 1989, \apjs, 71, 245

\bibitem[{{Kitchatinov} \& {R{\"u}diger}(1999)}]{1999A&A...344..911K}
{Kitchatinov}, L.~L. \& {R{\"u}diger}, G. 1999, \aap, 344, 911

\bibitem[{{Kochukhov}(2026)}]{kochukhov2026}
{Kochukhov}, O. 2026, \aap, 707, A16

\bibitem[{{Kochukhov} {et~al.}(2014){Kochukhov}, {L{\"u}ftinger}, {Neiner}, {Alecian}, \& {MiMeS Collaboration}}]{kochukhov2014}
{Kochukhov}, O., {L{\"u}ftinger}, T., {Neiner}, C., {Alecian}, E., \& {MiMeS Collaboration}. 2014, \aap, 565, A83

\bibitem[{{Kochukhov} {et~al.}(2010){Kochukhov}, {Makaganiuk}, \& {Piskunov}}]{kochukhov2010}
{Kochukhov}, O., {Makaganiuk}, V., \& {Piskunov}, N. 2010, \aap, 524, A5

\bibitem[{{Kriskovics} {et~al.}(2019){Kriskovics}, {K{\H{o}}v{\'a}ri}, {Vida}, {Ol{\'a}h}, {Carroll}, \& {Granzer}}]{kriskovics2019}
{Kriskovics}, L., {K{\H{o}}v{\'a}ri}, Z., {Vida}, K., {et~al.} 2019, \aap, 627, A52

\bibitem[{{K{\"u}ker} \& {R{\"u}diger}(1999)}]{1999A&A...346..922K}
{K{\"u}ker}, M. \& {R{\"u}diger}, G. 1999, \aap, 346, 922

\bibitem[{{Lachaume} {et~al.}(1999){Lachaume}, {Dominik}, {Lanz}, \& {Habing}}]{lachaume1999}
{Lachaume}, R., {Dominik}, C., {Lanz}, T., \& {Habing}, H.~J. 1999, \aap, 348, 897

\bibitem[{{Landi Degl'Innocenti} \& {Landolfi}(2004)}]{2004ASSL..307.....L}
{Landi Degl'Innocenti}, E. \& {Landolfi}, M. 2004, {Polarization in Spectral Lines}, Vol. 307

\bibitem[{{Luhman} {et~al.}(2007){Luhman}, {Patten}, {Marengo}, {Schuster}, {Hora}, {Ellis}, {Stauffer}, {Sonnett}, {Winston}, {Gutermuth}, {Megeath}, {Backman}, {Henry}, {Werner}, \& {Fazio}}]{luhman2007}
{Luhman}, K.~L., {Patten}, B.~M., {Marengo}, M., {et~al.} 2007, \apj, 654, 570

\bibitem[{{Mamajek} \& {Hillenbrand}(2008)}]{mamajek2008}
{Mamajek}, E.~E. \& {Hillenbrand}, L.~A. 2008, \apj, 687, 1264

\bibitem[{{Matt} {et~al.}(2012){Matt}, {MacGregor}, {Pinsonneault}, \& {Greene}}]{matt2012}
{Matt}, S.~P., {MacGregor}, K.~B., {Pinsonneault}, M.~H., \& {Greene}, T.~P. 2012, \apjl, 754, L26

\bibitem[{{McDonald} {et~al.}(2012){McDonald}, {Zijlstra}, \& {Boyer}}]{mcdonald2012}
{McDonald}, I., {Zijlstra}, A.~A., \& {Boyer}, M.~L. 2012, \mnras, 427, 343

\bibitem[{{Messina} \& {Guinan}(2002)}]{messina2002}
{Messina}, S. \& {Guinan}, E.~F. 2002, \aap, 393, 225

\bibitem[{{Messina} \& {Guinan}(2003)}]{messina2003}
{Messina}, S. \& {Guinan}, E.~F. 2003, \aap, 409, 1017

\bibitem[{{Parker}(1955)}]{parker1955}
{Parker}, E.~N. 1955, \apj, 122, 293

\bibitem[{{Perryman} {et~al.}(1998){Perryman}, {Brown}, {Lebreton}, {Gomez}, {Turon}, {Cayrel de Strobel}, {Mermilliod}, {Robichon}, {Kovalevsky}, \& {Crifo}}]{perryman1998}
{Perryman}, M.~A.~C., {Brown}, A.~G.~A., {Lebreton}, Y., {et~al.} 1998, \aap, 331, 81

\bibitem[{{Piskunov} {et~al.}(2011){Piskunov}, {Snik}, {Dolgopolov}, {Kochukhov}, {Rodenhuis}, {Valenti}, {Jeffers}, {Makaganiuk}, {Johns-Krull}, {Stempels}, \& {Keller}}]{piskunov2011}
{Piskunov}, N., {Snik}, F., {Dolgopolov}, A., {et~al.} 2011, The Messenger, 143, 7

\bibitem[{{Piskunov} \& {Valenti}(2002)}]{piskunov2002}
{Piskunov}, N.~E. \& {Valenti}, J.~A. 2002, \aap, 385, 1095

\bibitem[{{Ram{\'\i}rez} {et~al.}(2014){Ram{\'\i}rez}, {Mel{\'e}ndez}, \& {Asplund}}]{ramirez2014}
{Ram{\'\i}rez}, I., {Mel{\'e}ndez}, J., \& {Asplund}, M. 2014, \aap, 561, A7

\bibitem[{{R{\'e}ville} {et~al.}(2015){R{\'e}ville}, {Brun}, {Matt}, {Strugarek}, \& {Pinto}}]{reville2015}
{R{\'e}ville}, V., {Brun}, A.~S., {Matt}, S.~P., {Strugarek}, A., \& {Pinto}, R.~F. 2015, \apj, 798, 116

\bibitem[{{Ribas} {et~al.}(2010){Ribas}, {Porto de Mello}, {Ferreira}, {H{\'e}brard}, {Selsis}, {Catal{\'a}n}, {Garc{\'e}s}, {do Nascimento}, \& {de Medeiros}}]{ribas2010}
{Ribas}, I., {Porto de Mello}, G.~F., {Ferreira}, L.~D., {et~al.} 2010, \apj, 714, 384

\bibitem[{{Ros{\'e}n} {et~al.}(2016){Ros{\'e}n}, {Kochukhov}, {Hackman}, \& {Lehtinen}}]{rosen2016}
{Ros{\'e}n}, L., {Kochukhov}, O., {Hackman}, T., \& {Lehtinen}, J. 2016, \aap, 593, A35

\bibitem[{{Ryabchikova} {et~al.}(2015){Ryabchikova}, {Piskunov}, {Kurucz}, {Stempels}, {Heiter}, {Pakhomov}, \& {Barklem}}]{ryabchikova2015}
{Ryabchikova}, T., {Piskunov}, N., {Kurucz}, R.~L., {et~al.} 2015, \physscr, 90, 054005

\bibitem[{{Schr{\"o}der} {et~al.}(2013){Schr{\"o}der}, {Mittag}, {Hempelmann}, {Gonz{\'a}lez-P{\'e}rez}, \& {Schmitt}}]{schroder2013}
{Schr{\"o}der}, K.-P., {Mittag}, M., {Hempelmann}, A., {Gonz{\'a}lez-P{\'e}rez}, J.~N., \& {Schmitt}, J.~H.~M.~M. 2013, \aap, 554, A50

\bibitem[{{See} {et~al.}(2016){See}, {Jardine}, {Vidotto}, {Donati}, {Boro Saikia}, {Bouvier}, {Fares}, {Folsom}, {Gregory}, {Hussain}, {Jeffers}, {Marsden}, {Morin}, {Moutou}, {do Nascimento}, {Petit}, \& {Waite}}]{2016MNRAS.462.4442S}
{See}, V., {Jardine}, M., {Vidotto}, A.~A., {et~al.} 2016, \mnras, 462, 4442

\bibitem[{{See} {et~al.}(2014){See}, {Jardine}, {Vidotto}, {Petit}, {Marsden}, {Jeffers}, \& {do Nascimento}}]{see2014}
{See}, V., {Jardine}, M., {Vidotto}, A.~A., {et~al.} 2014, \aap, 570, A99

\bibitem[{{Semel}(1989)}]{semel1989}
{Semel}, M. 1989, \aap, 225, 456

\bibitem[{{Takeda} {et~al.}(2007){Takeda}, {Ford}, {Sills}, {Rasio}, {Fischer}, \& {Valenti}}]{takeda2007}
{Takeda}, G., {Ford}, E.~B., {Sills}, A., {et~al.} 2007, \apjs, 168, 297

\bibitem[{{Tuominen} {et~al.}(1999){Tuominen}, {Ilyin}, \& {Petrov}}]{tuominen1999}
{Tuominen}, I., {Ilyin}, I., \& {Petrov}, P. 1999, in Astrophysics with the NOT, ed. H.~{Karttunen} \& V.~{Piirola}, 47

\bibitem[{{Valenti} \& {Fischer}(2005)}]{valenti2005}
{Valenti}, J.~A. \& {Fischer}, D.~A. 2005, \apjs, 159, 141

\bibitem[{{Vican}(2012)}]{vican2012}
{Vican}, L. 2012, \aj, 143, 135

\bibitem[{{Vican} \& {Schneider}(2014)}]{vican2014}
{Vican}, L. \& {Schneider}, A. 2014, \apj, 780, 154

\bibitem[{{Viviani} {et~al.}(2018){Viviani}, {Warnecke}, {K{\"a}pyl{\"a}}, {K{\"a}pyl{\"a}}, {Olspert}, {Cole-Kodikara}, {Lehtinen}, \& {Brandenburg}}]{2018A&A...616A.160V}
{Viviani}, M., {Warnecke}, J., {K{\"a}pyl{\"a}}, M.~J., {et~al.} 2018, \aap, 616, A160

\bibitem[{{Willamo} {et~al.}(2022){Willamo}, {Lehtinen}, {Hackman}, {K{\"a}pyl{\"a}}, {Kochukhov}, {Jeffers}, {Korhonen}, \& {Marsden}}]{willamo2022}
{Willamo}, T., {Lehtinen}, J.~J., {Hackman}, T., {et~al.} 2022, \aap, 659, A71

\bibitem[{{Zuckerman} {et~al.}(2011){Zuckerman}, {Rhee}, {Song}, \& {Bessell}}]{zuckerman2011}
{Zuckerman}, B., {Rhee}, J.~H., {Song}, I., \& {Bessell}, M.~S. 2011, \apj, 732, 61

\end{thebibliography}

\begin{appendix}
\onecolumn
\section{List of observations \label{app:list_of_obs}}

\begin{table*}[htb!]
\centering
\caption{List of individual spectropolarimetric observations with HARPSpol. \label{tab:obs_log_harpspol}}
\begin{tabular}{lccrccccc}
\hline \hline
Date (UT) & HJD & $\phi_{\rm rot}$ & $S/N$ [V] & $\left\langle  B_z \right\rangle$ [G] & Null $\left\langle  B_z \right\rangle$ [G] & FAP [V] & FAP [null] & $\log R_{\rm HK}'$ \\
\hline
BE Cet & & & & & & & \\ \hline
2021 08 20 04:28 & 2459446.686 & 0.945 & 23317 & $-4.3 \pm 4.3$ & $-1.4 \pm 4.3$ & 0.02 & 0.92 & $-4.56$ \\
2021 08 21 05:08 & 2459447.714 & 0.079 & 100569 & $-2.4 \pm 1.0$ & $-0.7 \pm 1.0$ & $<10^{-16}$ & 0.93 & $-4.53$ \\
2021 08 22 05:22 & 2459448.724 & 0.211 & 59446 & $1.0 \pm 1.6$ & $1.4 \pm 1.6$ & $<10^{-16}$ & 0.90 & $-4.52$ \\
2021 08 27 06:38 & 2459453.777 & 0.869 & 78467 & $8.4 \pm 1.2$ & $0.5 \pm 1.2$ & $<10^{-16}$ & 0.93 & $-4.54$ \\
\hline
HN Peg & & & & & & & \\ \hline
2021 08 20 03:47 & 2459446.658 & 0.753 & 32951 & $-10.1 \pm 3.5$ & $-4.4 \pm 3.5$ & 0.02 & 0.86 & $-4.55$ \\
2021 08 21 03:20 & 2459447.639 & 0.966 & 74764 & $1.0 \pm 1.5$ & $2.9 \pm 1.5$ & $<10^{-16}$ & 0.98 & $-4.53$ \\
2021 08 21 06:34 & 2459447.774 & 0.995 & 74552 & $2.9 \pm 1.5$ & $-1.0 \pm 1.5$ & 1.89 $\times$ 10$^{-15}$ & 1.00 & $-4.53$ \\
2021 08 22 04:40 & 2459448.695 & 0.196 & 62473 & $-6.0 \pm 1.8$ & $0.8 \pm 1.8$ & $<10^{-16}$ & 0.99 & $-4.54$ \\
2021 08 27 02:54 & 2459453.621 & 0.267 & 80714 & $-10.8 \pm 1.4$ & $0.4 \pm 1.4$ & $<10^{-16}$ & 0.99 & $-4.52$ \\
2021 08 27 05:56 & 2459453.748 & 0.294 & 80733 & $-14.4 \pm 1.4$ & $1.9 \pm 1.4$ & $<10^{-16}$ & 0.96 & $-4.52$ \\
2021 08 28 02:15 & 2459454.594 & 0.478 & 75424 & $-4.0 \pm 1.5$ & $-0.0 \pm 1.5$ & $<10^{-16}$ & 1.00 & $-4.50$ \\
2021 08 28 05:14 & 2459454.719 & 0.505 & 80806 & $-4.9 \pm 1.4$ & $0.9 \pm 1.4$ & $<10^{-16}$ & 0.46 & $-4.50$ \\
2021 08 28 06:30 & 2459454.771 & 0.516 & 71980 & $-4.1 \pm 1.6$ & $-0.5 \pm 1.6$ & 1.67 $\times$ 10$^{-10}$ & 0.97 & $-4.50$ \\
\hline
$\kappa^1$~Cet & & & & & & & \\ \hline
2021 08 21 09:57 & 2459447.915 & 0.760 & 54771 & $-7.7 \pm 1.9$ & $0.2 \pm 1.9$ & $<10^{-16}$ & 1.00 & $-4.55$ \\
2021 08 22 06:10 & 2459448.757 & 0.852 & 55389 & $-9.0 \pm 1.9$ & $-0.8 \pm 1.9$ & $<10^{-16}$ & 1.00 & $-4.53$ \\
2021 08 27 07:06 & 2459453.796 & 0.399 & 80543 & $-11.7 \pm 1.3$ & $1.3 \pm 1.3$ & $<10^{-16}$ & 1.00 & $-4.57$ \\
2021 08 28 05:56 & 2459454.748 & 0.503 & 53928 & $-8.4 \pm 2.0$ & $0.7 \pm 2.0$ & $<10^{-16}$ & 0.99 & $-4.57$ \\
2021 08 28 09:31 & 2459454.897 & 0.519 & 81094 & $-5.6 \pm 1.3$ & $-0.2 \pm 1.3$ & $<10^{-16}$ & 1.00 & $-4.58$ \\
\hline
V1358~Ori & & & & & & & \\ \hline
2021 08 20 09:31 & 2459446.897 & 0.783 & 17615 & $-37.8 \pm 29.5$ & $-12.3 \pm 29.3$ & 0.96 & 1.00 & $-4.43$ \\
2021 08 21 09:18 & 2459447.888 & 0.513 & 27356 & $-12.9 \pm 18.5$ & $10.1 \pm 18.5$ & 0.70 & 1.00 & $-4.38$ \\
2021 08 22 09:31 & 2459448.897 & 0.256 & 23833 & $31.4 \pm 21.3$ & $0.0 \pm 21.2$ & 1.00 & 1.00 & $-4.38$ \\
2021 08 27 09:19 & 2459453.889 & 0.935 & 31854 & $-9.3 \pm 16.1$ & $11.1 \pm 16.0$ & 0.97 & 1.00 & $-4.38$ \\
2021 08 28 08:50 & 2459454.868 & 0.657 & 32781 & $20.7 \pm 15.5$ & $10.9 \pm 15.5$ & 0.63 & 1.00 & $-4.37$ \\
\hline
\end{tabular}
\tablefoot{
$\phi_{\rm rot}$ is the rotational phase. $S/N$ [V] is the signal-to-noise ratio of the Stokes~LSD~$V$ profile. $\left\langle  B_z \right\rangle$ is shown both for the Stokes~$V$ and the null~profile.}
\end{table*}

\begin{table}[htb!]
  \caption{List of individual spectropolarimetric observations of HN~Peg with SOFIN. \label{tab:obs_log_sofin}}
  \centering
\begin{tabular}{ccccccc}
  \hline \hline
Date (UT) & HJD & $\phi_{\rm rot}$ & $S/N$ & $\left\langle  B_z \right\rangle$ [G] & FAP [V] \\
\hline
2017 06 10 03:32 & 2457914.649 & 0.707 & 569 & $15.8 \pm 12.6$ & 0.87 \\
2017 06 11 03:46 & 2457915.659 & 0.927 & 518 & $24.9 \pm 10.5$ & 3.23 $\times$ 10$^{-3}$ \\
2017 06 12 03:36 & 2457916.652 & 0.142 & 453 & $8.7 \pm 10.8$ & 3.68 $\times$ 10$^{-3}$ \\
2017 06 13 03:53 & 2457917.664 & 0.363 & 430 & $-2.8 \pm 10.7$ & 0.33 \\
2017 06 14 03:29 & 2457918.647 & 0.576 & 448 & $-16.7 \pm 11.0$ & 0.49 \\
\hline
2021 07 17 04:06 & 2459412.675 & 0.365 & 430 & $-31.4 \pm 10.8$ & 1.74 $\times$ 10$^{-4}$ \\
2021 07 18 03:34 & 2459413.653 & 0.577 & 434 & $22.4 \pm 10.8$ & 7.47 $\times$ 10$^{-6}$ \\
2021 07 19 03:22 & 2459414.645 & 0.793 & 391 & $0.1 \pm 14.0$ & 0.10 \\
2021 07 20 02:20 & 2459415.602 & 0.001 & 372 & $-0.3 \pm 12.7$ & 0.13 \\
2021 07 21 03:21 & 2459416.644 & 0.227 & 544 & $29.2 \pm 9.0$ & 8.66 $\times$ 10$^{-7}$ \\
2021 07 22 02:52 & 2459417.624 & 0.440 & 435 & $11.5 \pm 10.7$ & 0.12 \\
2021 07 23 03:03 & 2459418.632 & 0.659 & 457 & $-2.4 \pm 14.1$ & 0.51 \\
2021 07 25 03:07 & 2459420.635 & 0.095 & 445 & $-42.0 \pm 11.7$ & 2.38 $\times$ 10$^{-10}$ \\
2021 07 26 04:05 & 2459421.675 & 0.321 & 426 & $20.1 \pm 11.8$ & 1.09 $\times$ 10$^{-4}$ \\
\hline
\hline
\end{tabular}
\tablefoot{
The $S/N$ signal-to-noise ratio is estimated for the Stokes $V$ spectra before LSD calculation.}
\label{obs}
\end{table}

\newpage
\section{Stokes~V profiles \label{app:stokes_v}}

\begin{figure}[htb!]
    \includegraphics[width=\textwidth]{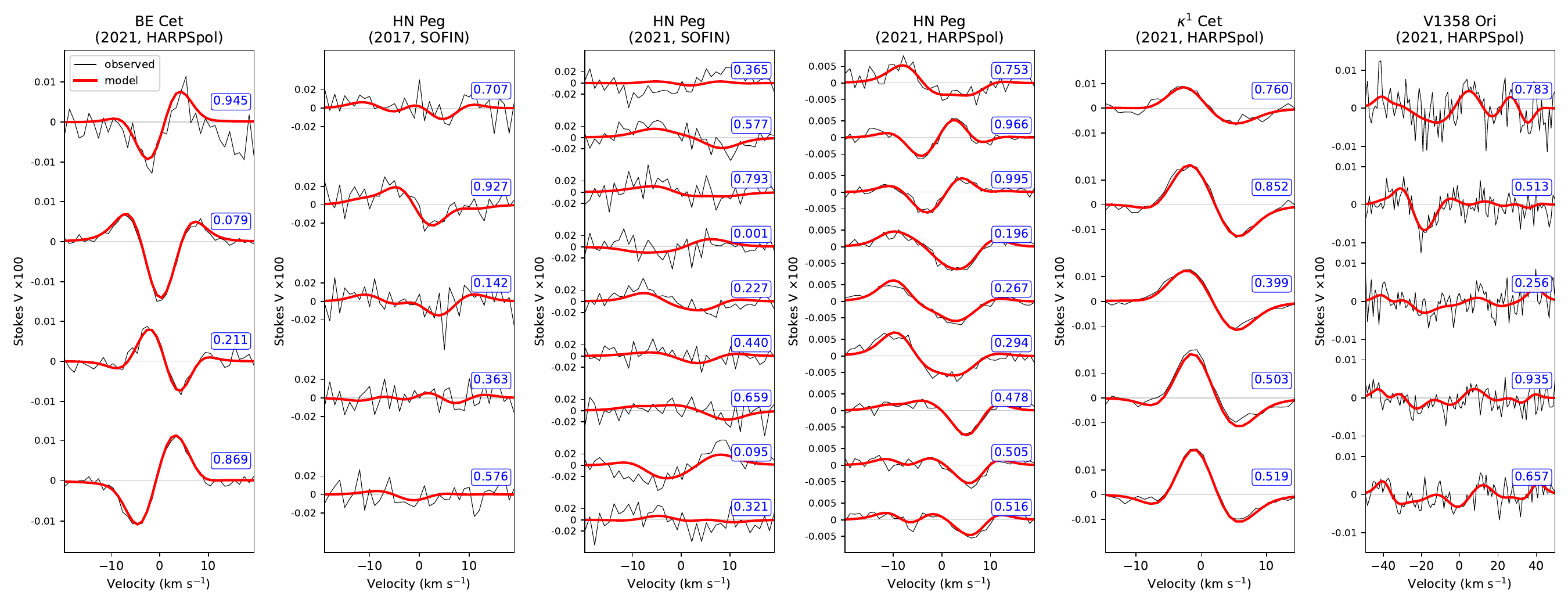}
      \caption{LSD profiles for BE~Cet, HN~Peg, $\kappa^1$~Cet, and V1358~Ori (from left to right, top to bottom). The year and the instrument of the observations are shown in parentheses. Observed Stokes~$V$ profiles are shown black, the best-fit solutions are shown in red. The phase of each observations is shown on the right side of each profile.
      }
      \label{fig:stokes_v_all}
\end{figure}

\section{Additional parameters of the ZDI maps \label{app:app_additional_params}}

\begin{table*}[htb!]
\centering
\caption{Additional parameters of the magnetic fields in the ZDI maps. \label{tab:app_additional_params}}
\begin{tabular}{lllcccccc}
\hline \hline
Star & Instrument & Epoch & $\left\langle |B| \right\rangle$ [G] & $|B|_{\rm max}$ [G] & Poloidal (\%) & Toroidal (\%) & Axisym.~(\%) & Non-axisym.~(\%) \\
\hline
BE Cet & HARPSpol & 2021.6 & 28.0 & 62.0 & 81.9 & 18.1 & 30.5 & 69.5  \\
HN Peg & SOFIN & 2017.4 & 24.1 & 74.3 & 72.8 & 27.2 & 57.6 & 42.4  \\
HN Peg & SOFIN & 2021.5 & 33.8 & 76.5 & 91.4 & 8.6 & 22.9 & 77.1  \\
HN Peg & HARPSpol & 2021.6 & 24.8 & 53.8 & 48.9 & 51.1 & 69.0 & 31.0  \\
$\kappa^1$~Cet & HARPSpol & 2021.6 & 34.5 & 79.4 & 17.5 & 82.5 & 87.7 & 12.3  \\
\hline
\end{tabular}
\end{table*}

\section{Fraction of magnetic energy in spherical harmonic expansions \label{app:fraction_E}}

\begin{table*}[htb!]
\centering
\caption{Fraction of magnetic energy (in \%) in each spherical harmonic expansion. \label{tab:app_fraction_E}}
\begin{tabular}{cccccccccccc}
\hline \hline
Star & $\ell = 1$ & $\ell = 2$ & $\ell = 3$ & $\ell = 4$ & $\ell = 5$ & $\ell = 6$ & $\ell = 7$ & $\ell = 8$ & $\ell = 9$ & $\ell = 10$ \\
\hline
BE Cet & 69.6 & 20.2 & 4.9 & 3.4 & 0.9 & 0.6 & 0.3 & 0.1 & 0.0 & 0.0 & \\
HN Peg (SOFIN, 2017) & 64.8 & 11.5 & 9.7 & 4.7 & 4.5 & 3.4 & 1.1 & 0.3 & 0.0 & 0.0 & \\
HN Peg (SOFIN, 2021) & 48.9 & 19.1 & 23.7 & 5.6 & 2.2 & 0.4 & 0.1 & 0.0 & 0.0 & 0.0 & \\
HN Peg & 64.4 & 8.1 & 19.5 & 4.1 & 2.2 & 1.0 & 0.4 & 0.2 & 0.1 & 0.0 & \\
$\kappa^1$ Cet & 72.6 & 5.0 & 14.5 & 5.4 & 1.4 & 0.6 & 0.4 & 0.1 & 0.0 & 0.0 & \\
\hline
\end{tabular}
\end{table*}

\end{appendix}
\end{document}